\documentclass[12pt]{article}
\usepackage[margin=1.1in]{geometry}
\usepackage[utf8]{inputenc}
\usepackage{amsthm,amsmath,physics,mathtools,amssymb}

\usepackage{charter}

\usepackage{CJKutf8}

\usepackage{enumitem}

\usepackage{hyperref}
\hypersetup{
    colorlinks=true,
    linkcolor=blue,
    filecolor=magenta,      
    urlcolor=cyan,
}
\usepackage[capitalise]{cleveref}

\theoremstyle{definition}

\usepackage{graphicx}
\graphicspath{ {images/} }

\usepackage{tensor}

\DeclarePairedDelimiterX{\inp}[2]{\langle}{\rangle}{#1, #2}

\usepackage{xparse}

\NewDocumentCommand\LH{mo}{%
  \IfNoValueTF{#2}
   {\mathcal{L}(\mathcal{H}^{#1})}
   {\mathcal{L}(\mathcal{H}^{#1},\mathcal{H}^{#2})}%
}

\newcommand\id{\leavevmode\hbox{\small1\kern-3.3pt\normalsize1}}

\allowdisplaybreaks

\usepackage{bbold}

\usepackage{authblk}

\usepackage[title]{appendix}

\usepackage{mciteplus}

\begin{document}

\begin{CJK*}{UTF8}{gbsn}
\title{What should be the ontology for the Standard Model?}
\author{Ding Jia (贾丁)\thanks{djia@perimeterinstitute.ca}}
\affil{Perimeter Institute for Theoretical Physics, Waterloo, Ontario, N2L 2Y5, Canada}
\affil{Department of Physics and Astronomy, University of Waterloo, Waterloo, Ontario, N2L 3G1, Canada}
\date{}
\maketitle
\end{CJK*}

\begin{abstract}
Although the Standard Model of particle physics is usually formulated in terms of fields, it can be equivalently formulated in terms of particles and strings. In this picture particles and open strings are always coupled. This offers an intuitive and graphical explanation for the otherwise mysterious gauge symmetry. In addition, the particle-string formulation avoids introducing redundant path integral configurations that are present in the field formulation. For its explanatory power and economy, the particle-string ontology may be preferred over the field ontology for the Standard Model.
\end{abstract}






\section{Introduction}

\begin{figure}
    \centering
    \includegraphics[width=.5\textwidth]{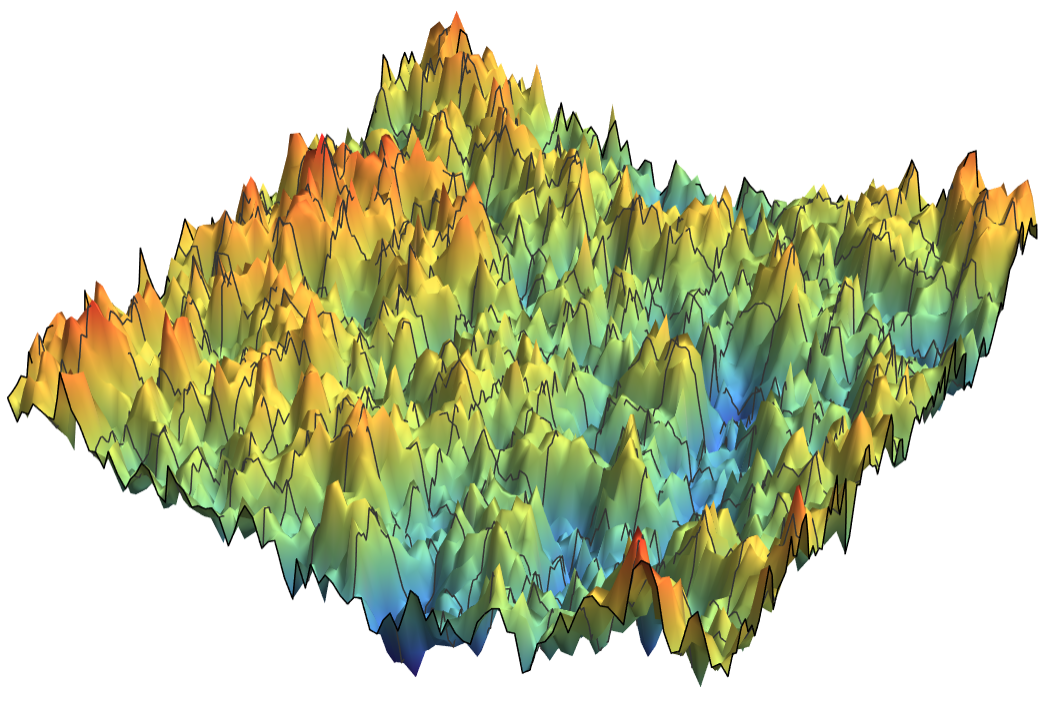}
    \caption{The field picture of the Standard Model: What exists are field configurations.}
    \label{fig:fc1}
\end{figure}

The Standard Model (of particle physics) is usually formulated as a quantum field theory. A simple-minded understanding of its ontology is in terms of fields. What exists in the Standard Model are field configurations put in superposition under the path integral (\cref{fig:fc1}).

\begin{figure}
    \centering
    \includegraphics[width=.8\textwidth]{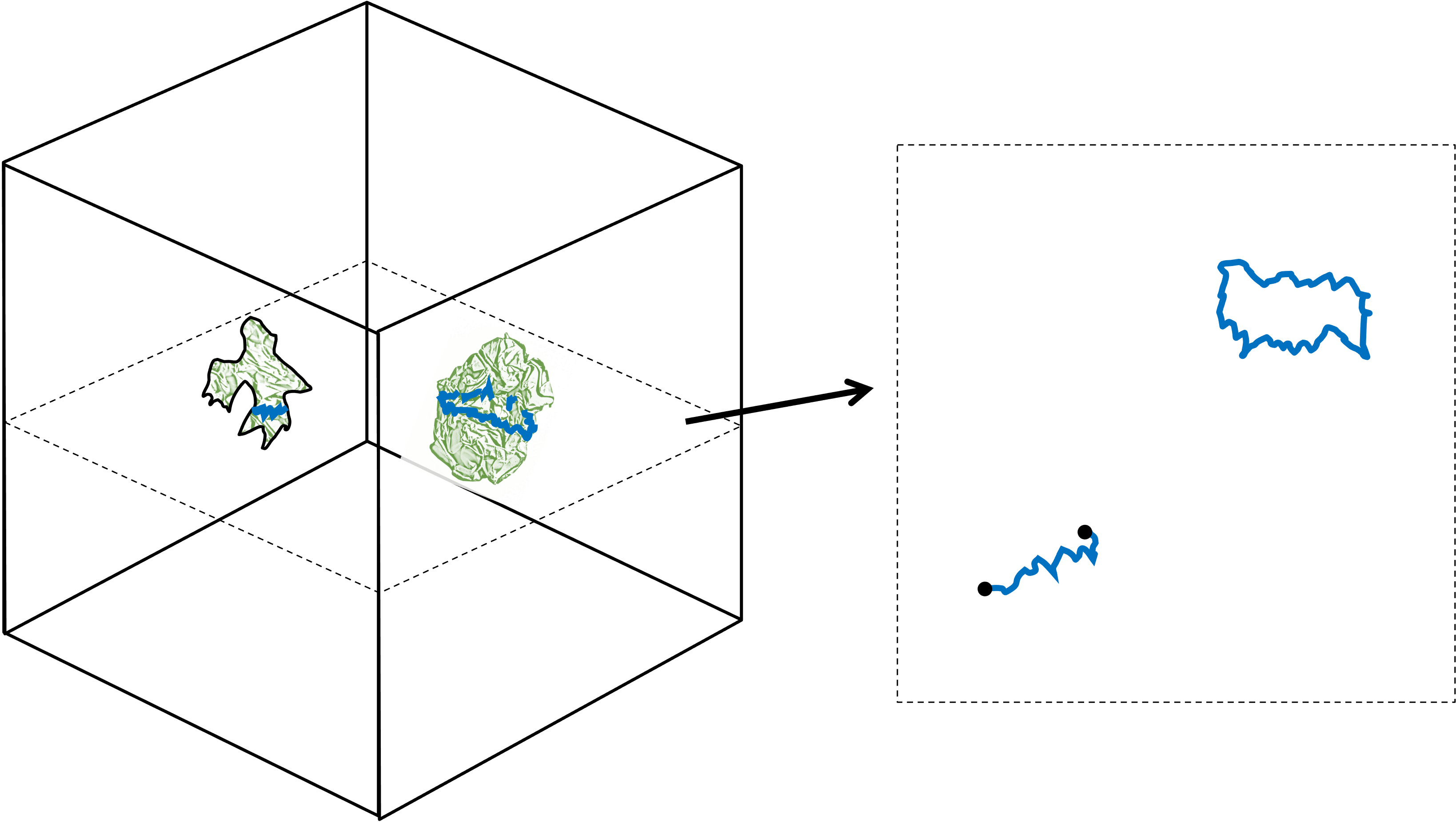}
    \caption{The particle-string picture of the Standard Model: What exists are particles and strings, which are generically non-smooth in path integral configurations. In spacetime (left figure), particles trace out $1D$ lines (thick black lines), while strings trace out $2D$ surfaces (green crumpled surfaces). In a hypersurface cross section (right figure), particle lines form points (black points), while string surfaces form lines (blue lines). Crucially, due to gauge symmetry, particles are always attached to strings, and strings are always either closed, or attached to particles.}
    \label{fig:psc1}
\end{figure}

However, there exists an equivalent reformulation of the Standard Model in terms of particles and strings.\footnote{This  particle-string formulation not directly related to Superstring theory. Neither extra spacetime dimensions nor supersymmetry is assumed.} In the particle-string formulation\footnote{Although it is known by many that quantum field theories can generically be re-expressed as theories of particles, strings, and higher-dimensional extended objects at perturbative and non-perturbative levels (see \cite{Feynman1950MathematicalInteraction, *Feynman1951AnElectrodynamics, Wilson1974ConfinementQuarks, Kleinert1989GaugeMatter, Fernandez1992RandomTheory, Bruegmann19924dMeasure, Ambjrn1997QuantumGeometry, Costello2011RenormalizationTheory, Gattringer2016ApproachesTheory, EdwardsQuantumTheory} and references therein), this piece of knowledge is not shared by the majority of the physicists. I will follow the works of Gattringer and collaborators \cite{Gattringer2013SpectroscopyGas, Mercado2013SurfaceLattice, Gattringer2016ApproachesTheory, Gattringer2018WorldlinesCycles, Marchis2018DualFluxes} to give a review of the particle-string reformulations at the non-perturbative level below. Such reformulations are originally proposed for technical motivations such as more efficient numerical computations. To my knowledge the conceptual implications have not been much investigated. This work is an attempt in this direction.}, what exists are particles and strings bounding each other and put in superposition under a path integral (\cref{fig:psc1}).

The two equivalent formulations pose a question. Is there a preferred ontological picture between the two? Should we think of fields or particles and strings as the basic entities for the Standard Model?


The main point of this work is to point out a conceptual reason to prefer the particle-string picture: It explains gauge symmetries. Suppose the particles are always attached to strings, and suppose the strings are either closed, or have their open ends attached to particles. Then gauge symmetry automatically holds, in the sense that a field reformulation of the particle-string theory automatically obeys gauge symmetries. \textit{In short, the reason for gauge symmetry is that particles and open strings are always coupled.}

In comparison, the field picture leaves gauge symmetries mysterious. There are certainly quantum field theories that are not gauge theories. Why gauge theories then? In the words of Rovelli \cite{Rovelli2014WhyGauge}:
\begin{quote}
Gauge theories are characterized by a local invariance, which is often described as mathematical redundancy. According to this interpretation, physics is coded into the gauge-invariant aspects of the mathematics. ... But things are not so clear. If gauge is only mathematical redundancy, why the common emphasis on the importance of gauge symmetry? Why the idea that this is a major discovery and guiding principle for understanding particle physics?

... Gauge theories are sometime introduced mentioning the historical idea of promoting a global symmetry to a local one. The purpose of the field would be to realize the local symmetry. This idea, however, leaves the question ... open: why do we need to describe the world with local symmetries if we then interpret these symmetries as mathematical redundancy?
\end{quote}
The lack of a convincing explanation for gauge symmetry in the field picture constitutes a reason to prefer the particle-string picture for the ontology of the Standard Model.

The particle-string picture is also more economic than the field picture. For theories with certain global or local symmetries, the field formulation sums over more path integral configurations than the particle-string formulation. These additional configurations cancel among themselves in the end in a field path integral. Avoiding this field redundancy from the outset leads to the particle-string formulation.

An additional motivation to consider the particle-string ontology comes from constructing new theories. For instance in discussions of quantum foundations, Wallace \cite{Wallace2020OnProblem} criticizes dynamical collapse models and Bohmian mechanics on the basis that it is much harder than is generally recognised to construct quantum field theory versions of them in order to incorporate the physical contents of the Standard Model. Part of the difficulty is that the dynamical collapse models and Bohmian mechanics studied in the context of non-relativistic quantum physics refer to particles but not fields. One might hope that the particle-string ontology for the Standard Model suggest ways to develop relativistic versions of dynamical collapse models and Bohmian mechanics without the need to migrate to a field ontology.

The above points are elaborated on below. In \cref{sec:sfp}, I review the particle reformulations of quantum field theories for matter fields. That $Z_2$ or $U(1)$ global symmetry holds is another way to say that particle lines must keep extending. In \cref{sec:gfs}, I review the string reformulations of quantum field theories for gauge fields. That $SU(N)$ local symmetry holds is another way to say that strings surfaces must keep extending. In \cref{sec:psc}, I review the particle-string reformulation of quantum field theories for matter-gauge coupled systems. That $SU(N)$ local symmetry holds is another way to say that particles and open strings are always coupled. In \cref{sec:fr}, I note that quantum field path integrals contain redundant configurations in the presence of the symmetries considered. The redundant configurations are avoided in the particle-string formalism. In \cref{sec:d}, I close with a discussion.

\section{Matter fields and particles}\label{sec:sfp}



\subsection{Real scalar field and unoriented particles}

Consider a real scalar field theory in Minkowski spacetime with the Lagrangian density
\begin{align}
    \mathcal{L}=-\frac{1}{2}\partial^\nu \phi \partial_\nu \phi-\frac{1}{2} m^2\phi^2(x) - V(\phi)
\end{align}
with a general potential $V$. To define the path integral non-perturbatively, a $D$-dimensional hypercubic lattice with spacing $a$ in both time and space directions is introduced. The lattice action reads
\begin{align}
S=&a^D \sum_x [ -\frac{1}{2} \sum_{\nu=1}^D g^{\nu\nu}(\frac{\phi_{x+\nu}-\phi_x}{a})^2 - \frac{1}{2} m^2 \phi_x^2 - V(\phi_x)]
\\
=& \sum_x [ \sum_{\nu=1}^D g^{\nu\nu} \tilde{\phi}_{x+\nu} \tilde{\phi}_{x} - \eta \tilde{\phi}_x^2  - \tilde{V}(\tilde{\phi}_x)],\label{eq:rsal3}
\end{align}
where $g^{\nu\nu}$ is the Minkowski metric, $x$ refers to lattice vertex, and $x+\nu$ refers to the vertex one unit in the $\nu$ direction away from $x$.
In the last line, $\tilde{\phi}_x:= a^{\frac{D-2}{2}}\phi_x$, $\eta:=a^2 m^2/2 + D-2$, and $\tilde{V}(\tilde{\phi}_x)=a^D V(\phi_{x})$. The tilde symbols are omitted in the following for simplicity.

\subsection*{Particle configurations arise from series expansion}

So far there are only field configurations. Particle configurations appear when the path integrand $e^{iS}$ is expressed in its series form. Let $S_{1}$ be the first term of (\ref{eq:rsal3}), $\prod_{x,\nu}:= \prod_x \prod_{\nu=1}^D$, and $\sum_{n}:=\prod_{x,\nu} \sum_{n_{x,\nu}=0}^\infty$. Then
\begin{align}
e^{iS_{1}}=&\prod_{x,\nu} \exp{ i g^{\nu\nu} \phi_{x+\nu} \phi_{x} }
=\sum_{n}\prod_{x,\nu} \frac{( i g^{\nu\nu}\phi_{x+\nu} \phi_{x})^{n_{x,\nu}} }{n_{x,\nu}!} 
\\
=&\sum_{n}
(\prod_{x,\nu}\frac{( i g^{\nu\nu})^{n_{x,\nu}}}{n_{x,\nu}! })
(\prod_{x} \phi_x^{\sum_{\nu=1}^D(n_{x,\nu}+n_{x-\nu,\nu})}),
\\
Z=\int D\phi ~ e^{iS}=& \sum_{n} (\prod_{x,\nu} \frac{( i g^{\nu\nu})^{n_{x,\nu}}}{n_{x,\nu}!}) (\prod_{x} \int_{-\infty}^\infty d\phi_x ~\phi_x^{n_x} e^{-i \eta \phi_x^2-i V(\phi_x)}) \label{eq:rsfz}
\\
=& \sum_{n} (\prod_{x,\nu} \frac{( i g^{\nu\nu})^{n_{x,\nu}}}{n_{x,\nu}!}) (\prod_{x} f(n_x)),
\label{eq:rsfz1}
\end{align}
where $f$ stands for the integral and $n_x:=\sum_{\nu=\pm 1}^{\pm D} n_{x,\nu}$.

\begin{figure}
    \centering
    \includegraphics[width=1.0\textwidth]{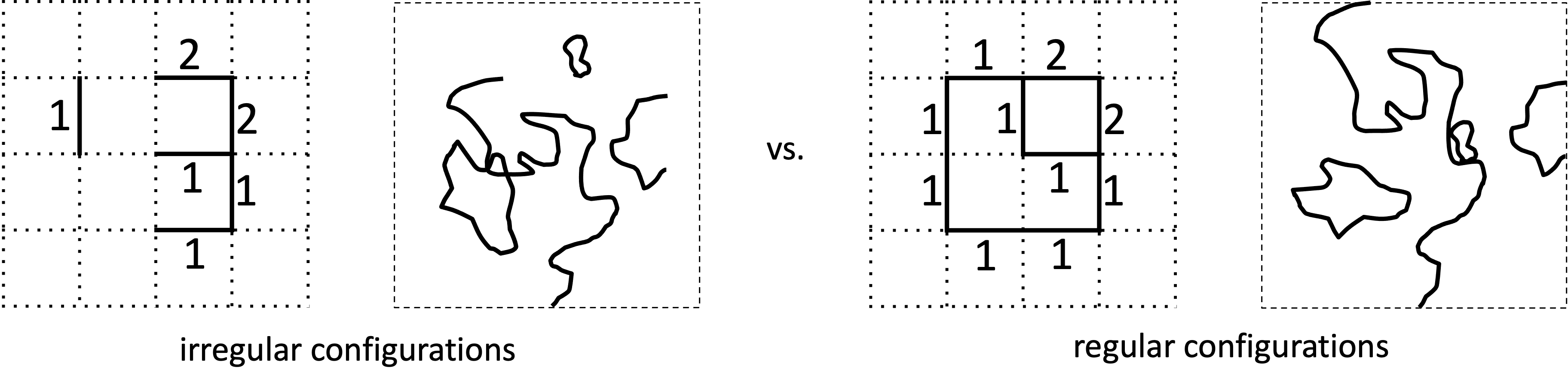}
    \caption{Left: lattice and continuum irregular configurations where particle lines do not need to extend (non-zero numbers of particle line segments are labelled on the lattice). Right: lattice and continuum regular configurations where particle lines keep extending.}
    \label{fig:ivelc1}
\end{figure}

These mathematically trivial steps are far from conceptually trivial. The basic entity of the theory has just changed from fields to particles in the following sense. We started with a path integral sum over field values $\phi$, but ended with a path integral sum over $n$ in \eqref{eq:rsfz1}. Here $n$ assigns a non-negative integer $n_{x,\nu}$ to each lattice edge $x,\nu$ connecting $x$ and $x+\nu$. Such an $n$-configuration admits an interpretation as a particle configuration, with $n_{x,\nu}$ as the number of particles passing the edge $x,\nu$, and $n_x$ as the total number of particle line segments passing $x$ (\cref{fig:ivelc1}). 
The path integral of \eqref{eq:rsfz1} is then understood as a sum over particle configurations.

\subsection*{Symmetry and extended particle lines}\label{sec:sepl}

From (\ref{eq:rsfz}),
\begin{align}
Z
=& \sum_{n}(\prod_{x,\nu} \frac{( i g^{\nu\nu})^{n_{x,\nu}}}{n_{x,\nu}!}) (\prod_{x} \int_0^\infty d r_x ~ r_x^{n_x}e^{-i\eta r_x^2}[e^{-iV(r_x)}+(-1)^{n_x}e^{-iV(-r_x)}] ), \label{eq:grsf}
\end{align}
where $\phi_x$ is expressed in terms of its magnitude $r_x:=\abs{\phi_x}$. When the theory obeys global $Z_2$ symmetry so that $V(r)=V(-r)$,
\begin{align}
Z =& \sum_{n}(\prod_{x,\nu} \frac{( i g^{\nu\nu})^{n_{x,\nu}}}{n_{x,\nu}!}) (\prod_{x} \int_0^\infty d r_x ~ r_x^{n_x}e^{-i\eta r_x^2}[(1+(-1)^{n_x})e^{-iV(r_x)}] )
\label{eq:rsfz2}
\\=& \sum_{n}(\prod_{x,\nu} \frac{( i g^{\nu\nu})^{n_{x,\nu}}}{n_{x,\nu}!}) (\prod_{x} 2 \delta_2(n_x)\int_0^\infty d r_x ~ r_x^{n_x}e^{-i\eta r_x^2-iV(r_x)}),
\end{align}
where $\delta_2(x)$ is the mod 2 Kronecker delta function. 

In symmetry considerations it is relevant to draw a distinction between regular and irregular configurations (\cref{fig:ivelc1}). A regular configuration is such that all the particle lines keep extending until they close on themselves or hit the boundary of the region of spacetime. This requires that in the interior of the region, each particle line segment that enters a vertex is paired with another particle line segment that exits the vertex. In terms of the particle numbers, this is ensured by requiring that at each vertex in the interior of the region, the integers at all the neighboring edges sum to an even value, i.e., by requiring $\delta_2(n_x)=1$ at all interior vertices $x$. On the other hand, a configuration that does not obey this requirement is considered irregular, and it contains particle lines that stop extending within the interior of the region.

Before any integration, the path integral (\ref{eq:rsfz}) and equivalently (\ref{eq:grsf}) include both irregular and regular configurations under $\sum_n$.

For a theory with global $Z_2$ symmetry, integrating the phase of $\phi$ (which is $\pm 1$ for a real scalar field) results in $\delta_2(n_x)$. All the irregular configurations cancel out among themselves in the phase sum to leave only the regular configurations where particle lines keep extending. In the particle formulation of the theory, we can declare that the path integral includes only the regular configurations from the outset.


For a theory without global $Z_2$ symmetry (e.g., with potential $V=\lambda_3 \phi^3+\lambda_4 \phi^4$), the irregular configurations are left over. Particle lines can pop up and disappear anywhere in spacetime. Therefore global $Z_2$ symmetry from the field perspective corresponds to the extendedness of particle lines from the particle perspective.

\subsection{Complex scalar field and oriented particles}\label{sec:csf}

Consider complex scalar field theories in Minkowski spacetime with the Lagrangian density
\begin{align}
    \mathcal{L}=-\partial^\nu \phi \partial_\nu \phi^*-m^2\abs{\phi}^2 - V(\phi).
\end{align}
The same steps of ``non-perturbative definition on lattice-series expansion-imposing symmetry'' leads to the following results which are straightforward to derive.

The lattice action reads
\begin{align}
S=& \sum_x [ \sum_{\nu=1}^D g^{\nu\nu}(\tilde{\phi}_x \tilde{\phi}^*_{x+\nu} + \tilde{\phi}_{x+\nu} \tilde{\phi}^*_{x} ) -\eta \abs{\tilde{\phi}_x}^2  - \tilde{V}(\tilde{\phi}_x)],\label{eq:csal3}
\end{align}
where $\tilde{\phi}_x:= a^{\frac{D-2}{2}}\phi_x$, $\eta:=a^2 m^2+2(D-2)$, and $\tilde{V}(\tilde{\phi}_x)=a^D V(\phi_{x})$. The tilde symbols are omitted in the following for simplicity.

Let $S_{1}$ be the first term in (\ref{eq:csal3}), $\prod_{x,\nu}:= \prod_x \prod_{\nu=1}^D$, and $\sum_{n}:=\prod_{x,\nu} \sum_{n_{x,\nu}=0}^\infty \sum_{n_{x+\nu,-\nu}=0}^\infty$. Like in the real scalar case, a series expansion leads to \cite{Gattringer2013SpectroscopyGas, Gattringer2016ApproachesTheory}
\begin{align}
e^{iS_{1}}=&\sum_{n}
(\prod_{x,\nu}\frac{(i g^{\nu\nu})^{n_{x,\nu}+n_{x+\nu,-\nu}}}{n_{x,\nu}! n_{x+\nu,-\nu}!})
(\prod_{x}\phi_x^{*\sum_\nu(n_{x,\nu}+n_{x,-\nu})} \phi_x^{\sum_\nu(n_{x+\nu,-\nu}+n_{x-\nu,\nu})}),
\\Z=&\sum_{n}(\prod_{x,\nu} \frac{(i g^{\nu\nu})^{n_{x,\nu}+n_{x+\nu,-\nu}}}{n_{x,\nu}! n_{x+\nu,-\nu}!}) (\prod_{x} \int_{-\pi}^{\pi} \frac{d\theta_x}{2\pi} e^{i \theta_x n_x} \int_0^\infty dr_x ~ r_x^{\bar{n}_x+1} e^{-i\eta r_x^2-iV(r_x e^{i\theta_x})}),\label{eq:csfgp}
\end{align}
where $\phi\in\mathbb{C}$ is expressed in polar form, and 
\begin{align}
n_x:=&\sum_{\nu=\pm 1}^{\pm D} (n_{x+\nu,-\nu}-n_{x,\nu}),\quad \bar{n}_x:=\sum_{\nu=\pm 1}^{\pm D}(n_{x+\nu,-\nu}+n_{x,\nu}).
\label{eq:nxo}
\end{align}

\begin{figure}
    \centering
    \includegraphics[width=1.0\textwidth]{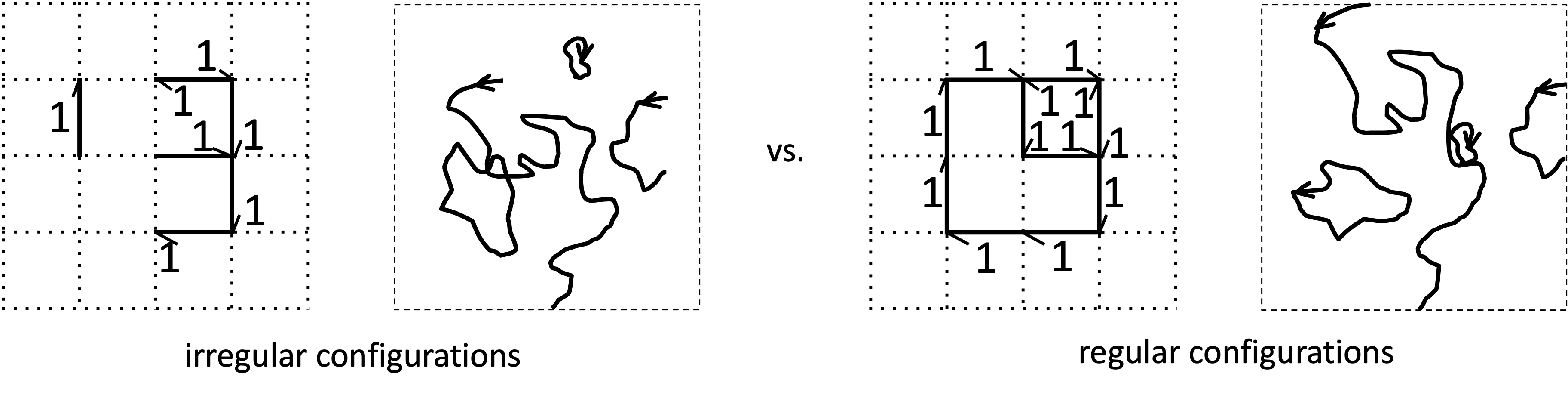}
    \caption{Left: lattice and continuum irregular configurations where oriented particle lines do not need to extend (non-zero numbers of particle line segments are labelled on the lattice). Right: lattice and continuum regular configurations where oriented particle lines keep extending.}
    \label{fig:ivelc3}
\end{figure} 

A lattice edge $x,\nu$ is associated with two particle numbers $n_{x,\nu}$ and $n_{x+\nu,-\nu}$ (\cref{fig:ivelc3}). In the case of real scalar field, we interpreted the one integer on an edge as the unoriented particle number. Here we interpret the two integers as the numbers of oriented particles travelling the edge in different directions. In particular, $n_{x,\nu}$ represents the number of particles travelling from $x$ in the positive $\nu$ direction, while $n_{x+\nu,-\nu}$ represents the number of particles travelling from $x+\nu$ in the negative $\nu$ direction. Then $n_x$ represents the difference between the numbers of particle line segments entering and exiting $x$, and $\bar{n}_x$ represents the total number of particle line segments crossing $x$.

Again, in symmetry considerations it is relevant to draw a distinction between regular and irregular configurations (\cref{fig:ivelc3}). A regular configuration is such that all the oriented particle lines keep extending until they close on themselves or hit the boundary of the region of spacetime. This requires that in the interior of the region, each particle line segment that enters a vertex is paired with another particle line segment that exits the vertex. Since $n_x$ represents the difference between the numbers of particle line segments entering and exiting $x$, this is ensured by requiring $\delta(n_x)=1$ at all interior vertices $x$. On the other hand, a configuration that does not obey this requirement is considered irregular, and it contains particle lines that stop extending within the interior of the region.

Before any integration, the path integral (\ref{eq:csfgp}) includes both irregular and regular configurations under $\sum_n$. 

For a theory with global $U(1)$ symmetry so that $V(r e^{i\theta})=V(r)$,
\begin{align}
Z=&\sum_{n}(\prod_{x,\nu} \frac{(i g^{\nu\nu})^{n_{x,\nu}+n_{x+\nu,-\nu}}}{n_{x,\nu}! n_{x+\nu,-\nu}!}) (\prod_{x} \delta(n_x) \int_0^\infty dr_x ~ r_x^{\bar{n}_x+1} e^{-i\eta r_x^2-iV(r_x)}).
\label{eq:csfu1s}
\end{align}
Integrating the phase of $\phi$ results in $\delta(n_x)$, which implies that the number of incoming and outgoing line segments are equal at all vertices. All the irregular configurations cancel out among themselves in the phase integral to leave only the regular configurations where particle lines keep extending (\cref{fig:ivelc3}). In the particle formulation of the theory, we can declare that the path integral includes only the regular configurations from the outset.

For a theory without global $U(1)$  symmetry (e.g., with potential $V=\lambda_3 \phi^3+\lambda_4 \abs{\phi}^4$), the irregular configurations are left over. Particle lines can pop up and disappear anywhere in spacetime. Therefore global $U(1)$ symmetry from the field perspective corresponds to the extendedness of oriented particle lines from the particle perspective.

\subsection{Fermion field and oriented particles}

For a fermionic theory in Minkowski spacetime with the Lagrangian density 
\begin{align}
\mathcal{L}=\overline{\psi} (i \gamma^\mu \partial_\mu -m)\psi,
\end{align}
the lattice action reads (after redefinitions to absorb constants)\footnote{In practical studies of lattice field theory one can adopt alternative actions such as with staggered fermions \cite{Gattringer2018WorldlinesCycles, Marchis2018DualFluxes} to ameliorate the fermion doubling problem. From a fundamental perspective perhaps it is more satisfactory to stick to the less \textit{ad hoc} naive fermion action and subject lattice lengths to path integration in quantum gravity \cite{Hamber2009QuantumApproach}.}
\begin{align}
S=& \sum_x [ \sum_{\mu=1}^D (\overline{\psi}_x i\gamma^\mu \psi_{x+\mu} - \overline{\psi}_{x+\mu} i\gamma^\mu \psi_{x}) - m \overline{\psi}_x \psi_x].
\label{eq:fla} 
\end{align}
With $\prod_x$ and $\prod_{x,\mu}$ as defined previously, a series expansion yields
\begin{align}
e^{iS}=& [\prod_x\sum_{s_x} (-im\overline{\psi}_x \psi_x)^{s_x}]
\prod_{x,\mu} [\sum_{n_{x,\mu}}(-\overline{\psi}_x \gamma^\mu \psi_{x+\mu})^{n_{x,\mu}} \sum_{n_{x+\mu,-\mu}}(\overline{\psi}_{x+\mu} \gamma^\mu \psi_{x})^{n_{x+\mu,-\mu}}],
\end{align}
where $s, n=0,1$ because Grassmann variables are nilpotent. For the partition function, Grassmann integration yields
\begin{align}
Z=& \int D[\overline{\psi},\psi] [\prod_x\sum_{s_x} (-im\overline{\psi}_x \psi_x)^{s_x}]
\prod_{x,\mu} [\sum_{n_{x,\mu}}(-\overline{\psi}_x \gamma^\mu \psi_{x+\mu})^{n_{x,\mu}} \sum_{n_{x+\mu,-\mu}}(\overline{\psi}_{x+\mu} \gamma^\mu \psi_{x})^{n_{x+\mu,-\mu}}]
\\
=& \sum_{s, n} (-im)^{s_x}  (-\gamma^\mu)^{n_{x,\mu}} (\gamma^\mu)^{n_{x+\mu,-\mu}} \prod_x \delta(u_x) \delta(v_x)
\\
=& C \sum_{n} (\frac{\gamma^\mu}{im})^{n_{x,\mu}} (-\frac{\gamma^\mu}{im})^{n_{x+\mu,-\mu}} \prod_x C_x(n).
\end{align}

In the second line, $u_x:=s_x+\sum_{\mu=\pm 1}^D n_{x,\mu}-1$, $v_x:=s_x+\sum_{\mu=\pm 1}^D n_{x+ \mu, -\mu}-1$, and the delta functions are induced by Grassmann integration. Pictorially, in each configuration a lattice site is either filled by a mass ``monomer'' counted by $s$, or is crossed by exactly one outgoing and exactly one incoming fermion line segment counted by $n$. 

In the third line a factor $-im$ is extracted from each site to form $C=(-im)^{N}$ where $N$ is the number of sites. For sites occupied by a monomer, this comes from the monomer factor. For sites on a fermion line, this induces a division by $-im$, which can be attributed to the $\gamma^\mu$ factors, since the number of sites on a fermion line equals the number of line segments on it (away from the boundary of the region). After the $s$-sum, the constraint $C_x(n)$ enforces that a site $x$ is crossed by either $0$ or $1$ fermion line. Explicitly
\begin{align}
C_x(n)=\delta(n_x)[\delta(\bar{n}_x)+\delta(\bar{n}_x-2)],
\end{align}
where $n_x$ and $\bar{n}_x$ are as defined in (\ref{eq:nxo}). They represent the difference in number for incoming and outgoing line segments, and the total number of line segments crossing $x$.

This picture of the extending oriented fermion particles is the same as that of the complex scalar field with $U(1)$ symmetry, except that no identical fermion line segments can overlap which enforces Pauli's exclusion principle. For fermions we do not consider theories with non-extending particle lines (irregular configurations) because any Lagrangian density where $\psi$ and $\overline{\psi}$ show up in pairs automatically enforces global $U(1)$ symmetry.

\section{Gauge fields and strings}\label{sec:gfs}

Quantum field theory presentations of gauge theories sometimes leave the impression that gauge matter is not much different from scalar and fermion matter. Gauge field, like scalar and fermion fields, is just another field, with perhaps more components. 

In contrast, the particle-string formulation makes it clear that gauge matter is a totally different species. While the scalar and spin-$1/2$ fermion matter are particles tracing out $1D$ lines in spacetime, gauge matter are strings tracing out $2D$ surfaces in spacetime. 

\subsection{Abelian gauge field and oriented surfaces}\label{sec:u1gt}

The standard way to define quantum gauge theories is through Wilson's lattice gauge theory formalism \cite{Wilson1974ConfinementQuarks}. For a $U(1)$ gauge field with the Lagrangian density $\mathcal{L}= \frac{\beta}{2} F_{\mu\nu}^2$, the lattice action in terms of group variables $U_{x,\mu}\in U(1)$ on edges is
\begin{align}
S= & \frac{\beta}{2} \sum_{x, \mu<\nu} g^{\mu\mu} g^{\nu\nu} (U_{x,\mu}U_{x+\mu,\nu} U_{x+\nu,\mu}^* U_{x,\nu}^* + U_{x,\mu}^* U_{x+\mu,\nu}^* U_{x+\nu,\mu} U_{x,\nu}),
\end{align}
where $\prod_{x,\mu<\nu}:= \prod_x \prod_{\nu=2}^D \prod_{\mu=1}^{\nu-1}$., which is a sum over all plaquettes (elementary surfaces) of the lattice.

\subsection*{String configurations arise from series expansion}
Let $I_n$ be the modified Bessel function defined by $e^{\frac{z}{2}(t+t^{-1})}=\sum_{n\in\mathbb{Z}} I_n(z) t^n$ for $z,t\in \mathbb{C}, t\ne 0$. Then \cite{Gattringer2016ApproachesTheory}
\begin{align}
Z =& \int D[U] e^{\frac{i\beta}{2} \sum_{x, \mu<\nu}  g^{\mu\mu} g^{\nu\nu} (U_{x,\mu}U_{x+\mu,\nu} U_{x+\nu,\mu}^* U_{x,\nu}^* + U_{x,\mu}^* U_{x+\mu,\nu}^* U_{x+\nu,\mu} U_{x,\nu})}
\label{eq:zu1g1}
\\
=& \int D[U] \prod_{x,\mu<\nu} \sum_{p_{x,\mu\nu}\in\mathbb{Z}} I_{p_{x,\mu\nu}}[i \beta g^{\mu\mu} g^{\nu\nu}] (U_{x,\mu}U_{x+\mu,\nu} U_{x+\nu,\mu}^* U_{x,\nu}^*)^{p_{x,\mu\nu}}
\\
=& \sum_{p} (\prod_{x,\mu<\nu} I_{p_{x,\mu\nu}}[i \beta g^{\mu\mu} g^{\nu\nu}]) \prod_{x,\mu} \int d U_{x,\mu} U_{x,\mu}^{p_{x,\mu}}
\label{eq:zu1g2}
\\
=& \sum_{p} (\prod_{x,\mu<\nu} I_{p_{x,\mu\nu}}[i \beta g^{\mu\mu} g^{\nu\nu}]) \prod_{x,\mu} \delta(p_{x,\mu}),
\label{eq:zu1g}
\end{align}
where $\sum_p:=\prod_{x,\mu<\nu} \sum_{p_{x,\mu\nu}\in\mathbb{Z}}$, $\prod_{x,\mu}:= \prod_x \prod_{\mu=1}^D$, and
\begin{align}\label{eq:pxmu}
p_{x,\mu}:=\sum_{\rho:\rho<\mu} (p_{x,\rho\mu}-p_{x-\rho,\rho\mu})-\sum_{\rho:\rho>\mu} (p_{x,\mu\rho}-p_{x-\rho,\mu\rho}).
\end{align}


We started with a path integral sum over field values $U$, but ended with a path integral sum over $p$ in \eqref{eq:zu1g}. Here $p$ assigns an integer $p_{x,\mu\nu}\in\mathbb{Z}$ to each lattice plaquette $x,\mu\nu$ starting at vertex $x$ and extending in directions $\mu$ and $\nu$. Such a $p$-configuration admits an interpretation as a string configuration, with $p_{x,\mu\nu}$ as the number of elementary surface segments at the lattice plaquette $x,\mu\nu$, positively or negatively orientated depending on the sign of $p_{x,\mu\nu}\in\mathbb{Z}$. The path integral of \eqref{eq:zu1g} is then understood as a sum over string configurations.

\subsection*{Symmetry and extended string surfaces}\label{sec:sess}

\begin{figure}
    \centering
    \includegraphics[width=0.6\textwidth]{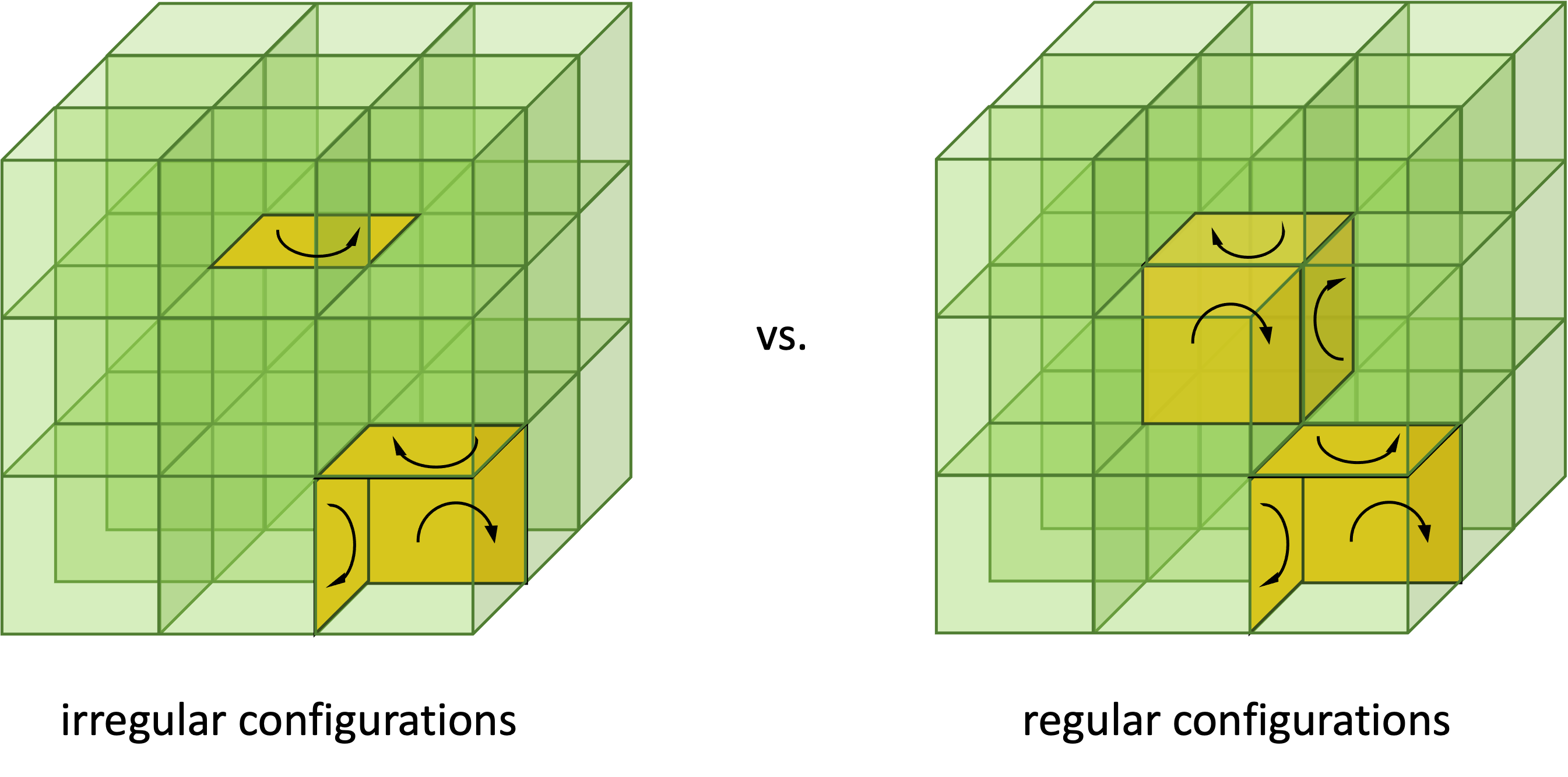}
    \caption{Left: an irregular surface configuration where positively and negatively oriented surface numbers do not match on some edges. Right: a regular surface configuration where positively and negatively oriented surface numbers match on all edges (in the interior of the region under consideration).}
    \label{fig:irsc2}
\end{figure} 

There is again a distinction between regular and irregular configurations (\cref{fig:irsc2}). A regular configuration is such that all the oriented surfaces keep extending until they close on themselves or hit the boundary of the region of spacetime. According to (\ref{eq:pxmu}), the net number (positive minus negative oriented) of surface segments crossing the edge $x,\mu$ is $p_{x,\mu}$. Requiring that the oriented surfaces always extend across the edge amounts to demanding $\delta(p_{x,\mu})=0$ at all interior edges, because this means all elementary surfaces touching this edge can be glued each other in a consistent orientation. On the other hand, a configuration that does not obey this requirement is considered irregular, and it contains string surfaces that stop extending within the interior of the region. 

Before any integration, the path integral \eqref{eq:zu1g2} include both irregular and regular configurations under $\sum_p$.

Integrating over $U$ in \eqref{eq:zu1g} gives rise to the delta function $\delta(p_{x,\mu})$, which ensure that the irregular configurations cancel out, and that only regular string configurations appear in the sum $\sum_p$. More generally, when a theory obeys gauge symmetry, the action contains only terms such as $U_{x,\mu}U_{x+\mu,\nu} U_{x+\nu,\mu}^* U_{x,\nu}^* $ where the edge variables $U$ form closed loops. A series expansion leads to a polynomial in the loops, which upon integration by $\int D[U]$ generates the delta function. In the surface picture, this implies that the oriented surfaces keep extending.

In contrast, suppose we start with a more general action $S= \int d^4 x [-\frac{\beta}{2} F_{\mu\nu}^2+V(A_\mu)]$, such as the Proca action with $V(A_\mu)=m^2 A_\mu A^\mu$. Then the non-perturbative theory has to resort from $U_{x,\mu}=e^{iaA_{x,\mu}}$ to $A_{x,\mu}$ as the basic variable in order to accommodate the additional term $V(A_\mu)$ in the action. The same procedure as in (\ref{eq:zu1g1}) to (\ref{eq:zu1g2}) yields
\begin{align}
Z =& \int D[A] e^{\frac{i\beta}{2} \sum_{x, \mu<\nu} g^{\mu\mu} g^{\nu\nu} (U_{x,\mu}U_{x+\mu,\nu} U_{x+\nu,\mu}^* U_{x,\nu}^* + U_{x,\mu}^* U_{x+\mu,\nu}^* U_{x+\nu,\mu} U_{x,\nu})} e^{i\sum_{x} V(A_{x,\mu})}
\\=& \sum_{p}(\prod_{x,\mu<\nu} I_{p_{x,\mu\nu}}[i \beta g^{\mu\mu} g^{\nu\nu}]) \int (\prod_{x,\mu} D A_{x,\mu}) ~e^{i a \sum_{x,\mu} A_{x,\mu} p_{x,\mu}+i\sum_{x} V(A_{x,\mu})}.
\label{eq:zu1ng}
\end{align}
For a general $V(A_{x,\mu})$ the constraint $\delta(p_{x,\mu})$ can no longer be derived. The surface picture where $\sum_p$ represents a sum over string configurations still holds, but the surfaces no longer need to keep extending. 

\subsection{Non-Abelian gauge field and colored oriented surfaces}\label{sec:nabgcos}

As an example of a non-Abelian gauge theory, consider $SU(3)$ gauge theory with the lattice action
\begin{align}
S= & \frac{\beta}{6} \sum_{x, \mu<\nu} \sum_{a,b,c,d=1}^3 g^{\mu\mu} g^{\nu\nu} (U_{x,\mu}^{ab} U_{x+\mu,\nu}^{bc} U_{x+\nu,\mu}^{dc~*} U_{x,\nu}^{ad~*} + U_{x,\mu}^{ab~*} U_{x+\mu,\nu}^{bc~*} U_{x+\nu,\mu}^{dc} U_{x,\nu}^{ad}),
\label{eq:su3a}
\end{align}
where $U^{ab}$ is the matrix representation of the $SU(3)$ group element in some basis. Similar to the $U(1)$ case, expanding the path integrand as a series in $U^{ab}$ and performing the group integration in terms of the various matrix components $U^{ab}$ yields delta function constraints. The explicit formulas which can be found in \cite{Gattringer2018WorldlinesCycles, Marchis2018DualFluxes} will not be shown here, because they are a bit lengthy and are not used below.

\begin{figure}
    \centering
    \includegraphics[width=1.0\textwidth]{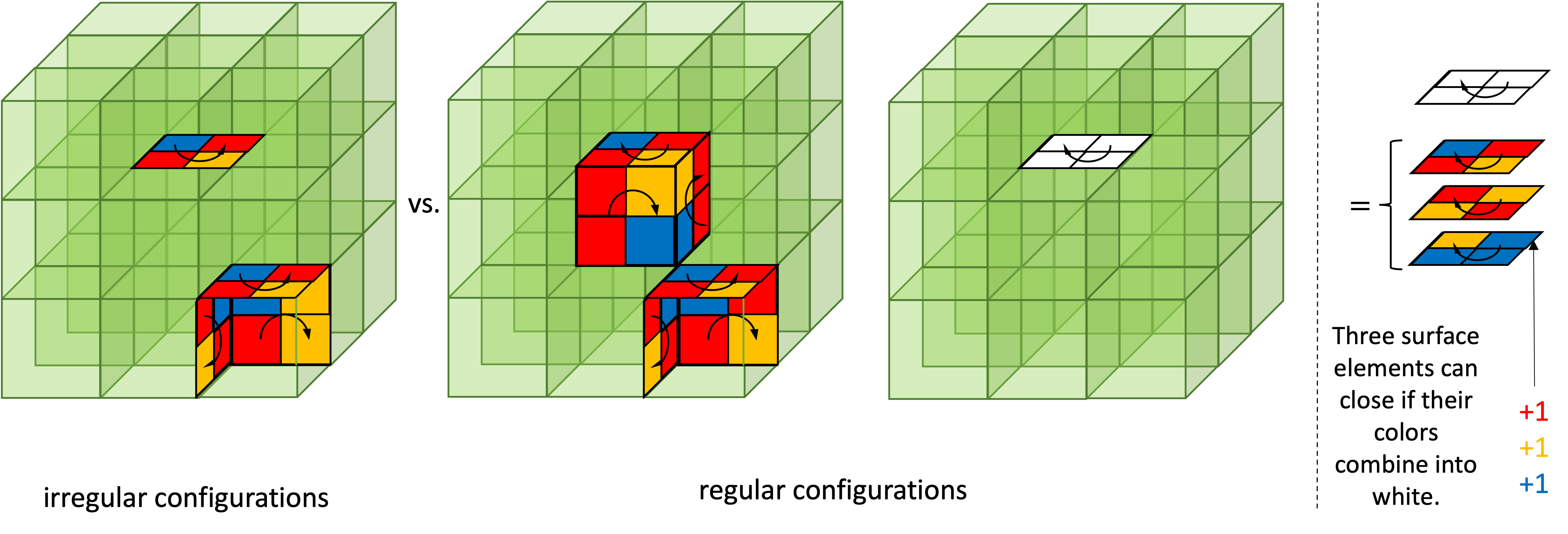}
    \caption{Left: an irregular surface configuration where either color or positively and negatively oriented surface numbers do not match on some edges. Middle: a regular surface configuration where both color and positively and negatively oriented surface numbers match on all edges (in the interior of the region under consideration). Right: a regular surface configuration closed by color combination.}
    \label{fig:ircsc}
\end{figure} 

Again, a picture of extended string surfaces  (\cref{fig:ircsc}) arises that captures the essence of the result. On each edge the $3\times 3$ matrix elements $U_{x,\mu}^{ab}$ introduces $3\times 3$ color combinations, e.g., red into yellow, red into blue etc. A lattice plaquette then has four color slots to support $3^4$ color combinations for a surface element. 

The constraints arising from group integration indicate that only regular configurations where surfaces keep extending stay (\cref{fig:ircsc}). Here a surface element extends by matching in color and cancelling in orientation (positive orientation cancels with negative orientation) on the common edge(s) with an adjacent surface element. This part is a straightforward generalization of the $U(1)$ case with one color. 

Interestingly, in $SU(3)$ surfaces can also close through color combination. Suppose three surface elements overlap in the same direction on an edge. At one of the two slots, if all three colors are present they combine into white. If the color is white at both slots of the edge, the three surfaces are considered to have no boundary at this edge. Through color combination, new closed surfaces can form, such as the totally white surface on the right of \cref{fig:ircsc}.

Again, the constraints are generated out of group integration. If we start with a theory without non-Abelian gauge symmetry, the constraints can no longer be derived. Pictorially, colored oriented surfaces no longer need to keep extending.

\section{Particle-string coupling}\label{sec:psc}

Consider a theory with both particles and strings. Suppose the theory obeys gauge symmetry (in the field reformulation). Then as demonstrated below, the particles are always attached to the strings, and the strings are either closed, or have their open ends attached to the particles. This need not hold if the theory does not obey gauge symmetry. 

If one adopts the particle-string ontology, then an intuitive explanation for gauge symmetry is available. Suppose the particles and open strings are always coupled. Then the field reformulation for the theory automatically obeys gauge symmetry. In this view, gauge symmetry is no longer a guiding principle, but only a derived property.

\subsection{Particles and uncolored strings}

Consider a $U(1)$-locally symmetric scalar-gauge coupled theory with the Lagrangian density
\begin{align}
    \mathcal{L}=-\abs{D_\mu \phi}^2-m^2\abs{\phi}^2+V(\phi_x)+\frac{\beta}{2} F_{\mu\nu}^2.
\end{align}
where $D_\mu=\partial_\mu+igA_\mu$ and $V$ is a $U(1)$-locally symmetric potential. Non-perturbatively, the scalar part action changes to
\begin{align}
S_P=& \sum_x [ \sum_{\nu=1}^D g^{\nu\nu}(\phi_x U_{x,\nu}^* \phi^*_{x+\nu} + \phi_{x+\nu} U_{x,\nu} \phi^*_{x} ) -\eta \abs{\phi_x}^2  - V(\phi_x)],\label{eq:csga}
\end{align}
where suitable redefinitions are performed as in \cref{sec:csf}. In comparison to (\ref{eq:csal3}), the only difference is that every near neighbor coupling $\phi_x  \phi^*_{x+\nu}$ is now dressed with $U_{x,\nu}^*$. After the $\phi$ integration, the same steps as in \cref{sec:csf} leads to
\begin{align}\label{eq:ZP}
Z_P=&\sum_{n}(\prod_{x,\nu} \frac{(i g^{\nu\nu}U_{x,\nu} )^{n_{x,\nu}} (ig^{\nu\nu}U_{x,\nu}^* )^{n_{x+\nu,-\nu}}}{n_{x,\nu}! n_{x+\nu,-\nu}!}) (\prod_{x} \delta(n_x) \int_0^\infty dr_x ~ r_x^{\bar{n}_x+1} e^{-i\eta r_x^2-iV(r_x)})
\end{align}
for the particle part. As in \cref{sec:csf}, we interpret the $n$-configurations as oriented particle configurations. Due to the delta function, the extended particle lines picture persists for the particle part of this $U(1)$-locally symmetric theory.

The gauge part of the theory is the same as in \cref{sec:u1gt}, except for the $U$ factors coming from  the dresses on the particle lines. Therefore after the $U$ integration, (\ref{eq:zu1g}) is replaced by
\begin{align}
\sum_{p} (\prod_{x,\mu<\nu} I_{p_{x,\mu\nu}}[i \beta g^{\mu\mu} g^{\nu\nu}]) \prod_{x,\mu} \delta(p_{x,\mu}+n_{x,\mu}).
\label{eq:zsu1g}
\end{align}
All together,
\begin{align}
Z=&\sum_{p} \sum_{n}(\prod_{x,\nu} \frac{(i g^{\nu\nu} )^{n_{x,\nu}+n_{x+\nu,-\nu}}}{n_{x,\nu}! n_{x+\nu,-\nu}!}) (\prod_{x} \delta(n_x) \int_0^\infty dr_x ~ r_x^{\bar{n}_x+1} e^{-i\eta r_x^2-iV(r_x)})+
\nonumber
\\&(\prod_{x,\mu<\nu} I_{p_{x,\mu\nu}}[i \beta g^{\mu\mu} g^{\nu\nu}]) \prod_{x,\mu} \delta(p_{x,\mu}+n_{x,\mu}).
\label{eq:Zps}
\end{align}
As in \cref{sec:u1gt}, we interpret the $p$-configurations as oriented string configurations, with $p_{x,\mu\nu}\in\mathbb{Z}$ as the number of elementary surface segments at the lattice plaquette $x,\mu\nu$, positively or negatively orientated depending on the sign of $p_{x,\mu\nu}$.

\begin{figure}
    \centering
    \includegraphics[width=0.75\textwidth]{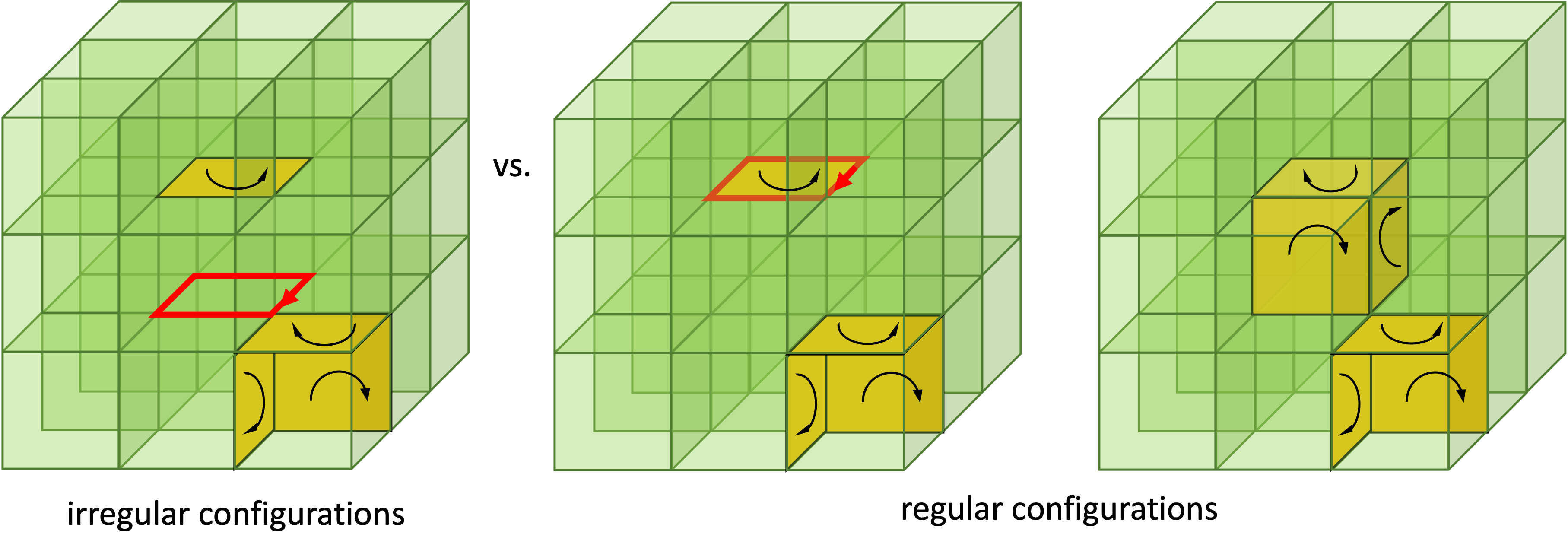}
    \caption{Left: an irregular line-surface configuration where positively and negatively oriented line and surface numbers do not match on some edges. Right: regular surface configurations where positively and negatively oriented line and surface numbers match on all edges (in the interior of the region under consideration).}
    \label{fig:irlsc}
\end{figure} 

The delta function $\delta(p_{x,\mu}+n_{x,\mu})$ of \eqref{eq:Zps} indicates that the number of positively or negatively oriented elementary particle lines matches with the number of negatively or positively oriented elementary surfaces at the edges $x,\mu$. Pictorially, this admits the interpretation that particle lines are always attached to string surfaces, and the open boundaries of string surfaces are always bounded by particle lines. In addition, the orientations are always matched (\cref{fig:irlsc}). Referring to configurations obeying these conditions as regular configurations and those that do not as irregular configurations, we can declare that in the particle-string formulation of the theory, the path integral defined by \eqref{eq:Zps} includes only the regular configurations from the outset.

If we started with a theory without gauge symmetry, this picture no longer holds. For instance, suppose
\begin{align}
V(\phi_x)=\sum_{\nu=1}^D g^{\nu\nu}(\phi_x \phi^*_{x+\nu} + \phi_{x+\nu} \phi^*_{x})
\end{align}
is as in a bare scalar action without gauge coupling. Then in place of \eqref{eq:csga} and \eqref{eq:ZP} we would get
\begin{align}
S_P=& \sum_x [ \sum_{\nu=1}^D g^{\nu\nu}[\phi_x (U_{x,\nu}^*+1) \phi^*_{x+\nu} + \phi_{x+\nu} (U_{x,\nu}+1) \phi^*_{x} ] -\eta \abs{\phi_x}^2],
\\
Z_P=&\sum_{n}(\prod_{x,\nu} \frac{[i g^{\nu\nu}(U_{x,\nu}+1) ]^{n_{x,\nu}} [ig^{\nu\nu}(U_{x,\nu}^*+1)]^{n_{x+\nu,-\nu}}}{n_{x,\nu}! n_{x+\nu,-\nu}!}) (\prod_{x} \delta(n_x) \int_0^\infty dr_x ~ r_x^{\bar{n}_x+1} e^{-i\eta r_x^2}).
\end{align}
Since $U$ now shows up as $(U_{x,\nu}^*+1)$ and $(U_{x,\nu}^*+1)$,the integrals over $U$ on the edges would no longer yield delta functions $\delta(p_{x,\mu}+n_{x,\mu})$ on the edges. This non-gauge invariant theory therefore include irregular configurations where particle lines and open string surfaces are not attached to each other. This example shows why the gauge-invariant coupling
\begin{align}
    g^{\nu\nu}(\phi_x U_{x,\nu}^* \phi^*_{x+\nu} + \phi_{x+\nu} U_{x,\nu} \phi^*_{x} )
\end{align}
for $\phi$ on adjacent vertices is crucial for generating the delta functions $\delta(p_{x,\mu}+n_{x,\mu})$ on the edges to exclude irregular configurations.

For fermion particles coupled to strings, apart from the fact that identical fermionic lines segments cannot overlap (Pauli's exclusion principle), the picture is the same.

\subsection{Particles and colored strings}

For a non-Abelian locally symmetric scalar-gauge coupled theory, the gauge part of the action is as (\ref{eq:su3a}), and the scalar part of the action is
\begin{align}
S_P=& \sum_x [ \sum_{\nu=1}^D \sum_{a,b} g^{\nu\nu}(\phi_{x}^a U_{x,\nu}^{ab~*} \phi^{b~*}_{x+\nu} + \phi_{x+\nu}^a U_{x,\nu}^{ab} \phi^{b~*}_{x} ) -\eta \abs{\phi_x^a}^2  - V(\phi_x^a)],\label{eq:csnaga}
\end{align}
where the sum $\sum_a$ is over colors. 

The same steps as in the last example leads to a picture of colored particles coupled to colored strings. Since a particle line segment is now dressed with $U^{ab}$, it also carries two colors. Group integration implies that the colors of the particle lines and of the string surfaces must cancel. In addition to the closed string surfaces shown in \cref{sec:nabgcos}, there are now open string surfaces bounded by particle lines, whose colors and orientations match. 

Again, the picture of the fermion particles \cite{Gattringer2018WorldlinesCycles, Marchis2018DualFluxes} is quite the same apart from the fact that identical fermionic lines cannot overlap (Pauli's exclusion principle).

\section{Field redundancy and partial local symmetry}\label{sec:fr}

\subsection{Field redundancy}

In the previous examples, quantum field theories with symmetries exhibit a redundancy. The irregular configurations cancel out among all themselves in the path integral sum. Only the regular configurations need to be included from the very beginning. 

In terms of particles and strings, the path integral can be defined to include only regular configurations. The redundancy is avoided. 

In contrast, in terms of fields the irregular configurations seems unavoidable because the form of quantum field theories is tightly constrained \cite{Weinberg1995TheFieldsb}. The particle-string formalism is more economic than the field formalism in this regard.

Because this form of redundancy is attached to the field formalism and can be avoided in the particle-string formalism, I call it \textbf{field redundancy}. Field redundancy is distinct from the gauge redundancy that relates gauge equivalent configurations. In the previous examples, field redundancy is seen for a broader class of symmetries including discrete, continuous, local, global, Abelian, and non-Abelian symmetries. 

Field redundancy is also a quantum property foreign to classical theory, because the cancellation is among configurations in superposition under a path integral.


\subsection{Partial local symmetries}


As an aside question, is there a more precise way to capture the relation of field redundancy to symmetry?

It is tempting to understand the cancellation in terms of Noether's theorem that relates charge conservation to symmetry, because particles and strings that keep extending seem to suggest some form of conservation law. However, Noether's theorem cannot be the answer. Firstly, Noether's theorem does not cover discrete symmetries such as the $Z_2$ symmetry of the real scalar field. Secondly, non-Abelian gauge theories do not have gauge-invariant Noether currents \cite{Schwartz2013QuantumModel}, whereas the regular particle-string configurations here are gauge invariant configurations.

The true answer seems to be given by what I call \textbf{partial local symmetry}. Consider path integrals that can be re-expressed as follows.
\begin{align*}
Z=&\int D[Y] ~A(Y)
\\
=&\int D[R] (\prod_x D[G_x]) ~A(R,\{G_x\})
\\
=&\int D[R] ~A_0(R) \prod_x \underbrace{(\int D[G_x]~ P_x(R,G_x))}_{C_x}.
\end{align*}
In the second line, the original variable $Y$ is decomposed into group variables $\{G_x\}$ and residue variable(s) $R$, and the amplitude $A(Y)$ is re-expressed in these new variables as $A(R,\{G_x\})$. Here the group variables $G_x$ are located to places $x$, which can be lattice points, edges, plaquettes etc.

In the third line, $A(R,\{G_x\}) = A_0(R) (\prod_x A_x(R,G_x))$ decomposes into two parts. The $A_0(R)$ part is independent of the group variables so can be taken out of the group integrals. This part is invariant under the local group actions by $G_x$ at any location $x$. Therefore the theory exhibits a form of local symmetry. Since $A_0(R)$ is only part of the whole amplitude $A(Y)$, the symmetry is dubbed ``partial local symmetry''.

The other part $P_x(R,G_x)$ are polynomials in the group variables $G_x$. Group integration $\int D[G_x]~ A_x(R,G_x)$ generates local constraints $C_x$ for the residue variables $R$.

For example, for the $Z_2$-symmetric real scalar field theory (\ref{eq:rsfz2}), $C_x = \sum_{z=\pm 1} z^{n_x}=\delta_2(n_x)$ with $G_x\in Z_2$, $R=n_x$, and $P_x=G_x^{n_x}$. As another example, for the $U(1)$-symmetric complex scalar field theory (\ref{eq:csfu1s}), $C_x = \int_{-\pi}^{\pi} \frac{d\theta_x}{2\pi} e^{i \theta_x n_x}=\delta(n_x)$ with $G_x\in U(1)$, $R=n_x$, and $P_x=G_x^{n_x}$. The other examples in the previous sections also follow the same pattern. 

Although these theories do not necessarily exhibit local gauge symmetry, they do exhibit partial local symmetry. The constraints that arise result in cancellations among field configurations, and hence imply field redundancy.



\section{Discussion}\label{sec:d}

Because the Standard Model is commonly formulated as a quantum field theory, it is tempting to consider fields as its basic entity, i.e., its ontology. In light of the particle-string reformulation reviewed here\footnote{To be precise, the particle-string formulation of the  full Standard Model is not explicitly given, but it can be straightforwardly obtained by generalizing the particle-string formulation of the matter-gauge coupled theories.}, the question about the ontology of the Standard Model deserves a deeper thought.

In the particle-string formulation, particles and open strings are always coupled. This property gives an intuitive explanation for the otherwise mysterious gauge symmetry of the field formulation. In addition, the field formulation includes redundant configurations that eventually cancel out among themselves. These redundant configurations are avoided from the outset in the particle-string formulation. For its explanatory power and its economy, the particle-string ontology may be preferred over the field ontology for the Standard Model.

One could question if the particle-string formulation really explains gauge symmetry better. After all, \textit{a priori} particles and open strings need not always be coupled, and one still needs to assume that they are in order to explain gauge symmetry. If it costs an extra assumption anyways, why could we not simply assume that the gauge fields obey a local symmetry to explain gauge symmetry?

I believe the particle-string explanation is still preferable. Imagine we are to explain the fundamental properties of matter to school students or laypeople eager for the scientific knowledge. The gauge symmetry principle is such an profound property that we do not want to miss. One way to explain it is: ``Matter are made of fields. The fields can be transformed according to local group actions. In our universe, the dynamical laws for matter are unchanged under such transformations.'' Another way is: ``Matter are made of particles and strings. In our universe, particles and open strings are always coupled.'' Which one do you prefer? An explanation is supposed to build intuition. The former ``explanation'' hardly builds any intuition at all, and should perhaps better be characterized as a ``description'' instead of an ``explanation'' for gauge symmetry. The latter explanation reduces the complicated mathematical concept of gauge symmetry to an easily visualizable picture of coupled particle lines and string surfaces. It does help build intuition, and is in this sense preferable.

These in no way implies that we should abandon the field formulation. A calculation that is hard in one formulation can be easier in another. For practical uses it is better to have more formulations in our toolbox than fewer, even though for conceptual understandings one formulation may be preferred.


In addition to these, reformulating a known theory in another picture can suggest different ideas towards discovering the unknowns. It is worth exploring ideas of beyond the Standard Model coming from the particle-string picture. For instance, a $4D$ space time has room to support higher dimensional objects in addition to $1D$ particle lines and $2D$ string surfaces. Could dark matter be such higher dimensional objects?


\section*{Acknowledgement}

I thank Christof Gattringer and Carlo Rovelli for helpful correspondences. I am very grateful to Lucien Hardy and Achim Kempf for long-term support and encouragement. 

Research at Perimeter Institute is supported in part by the Government of Canada through the Department of Innovation, Science and Economic Development Canada and by the Province of Ontario through the Ministry of Economic Development, Job Creation and Trade.

\bibliographystyle{unsrt}
\bibliography{mendeley.bib}
\end{document}